\documentclass[epsf,twocolumn,showpacs,preprintnumbers]{revtex4}
\usepackage{amsmath,amssymb,amsfonts}
\usepackage{graphics}
\usepackage{dcolumn} % Align table columns on decimal point
\usepackage{bm,graphicx,color}
\usepackage{epsfig,comment}
\usepackage{multirow}

\usepackage{comment}
\usepackage{graphicx}
\usepackage{dcolumn}
\usepackage{bm}
\usepackage{color}

\begin{document} 
\title{Why Compressed Metal Hydrides are \\ Near-room-temperature Superconductors}
\author{Warren E. Pickett}
%\thanks{\email{wepickett@ucdavis.edu}}\\
\affiliation{University of California Davis, Davis CA 95616}
\email{wepickett@ucdavis.edu}
\date{\today}
%\baselineskip24pt
%\maketitle 
%\begin{multicols*}{1}
\begin{abstract}
This contribution provides a partial
response to the titular statement since, it will be claimed, 
the ``why'' is not yet understood, but there is
a pathway for achieving a more complete understanding. 
The sense of the community has been that, given a prospective 
metal hydride and pressure, the energy landscape can be surveyed computationally for 
thermodynamic and dynamic stability, 
the Eliashberg spectral function with its required input (energy bands, phonon modes, coupling
matrix elements) 
can be calculated, and the critical temperature T$_c$ obtained. Satisfyingly large values of 
the electron-phonon coupling strength $\lambda$=2-3 at high mean frequency are obtained,
giving very reasonable agreement with  existing high T$_c$ hydrides. Typically 80-85\% of
$\lambda$ is
attributable to high frequency H vibrations. This much was envisioned by Ashcroft 
two decades ago, so why should there be any angst? This paper addresses more 
specifically the question {\it why hydrogen?} Light mass is indeed a factor, but
with possibilities not yet explored. 
This paper provides a concise overview of related formal developments occurring 
sporadically over several decades
that, when implemented, could resolve the question of
{\it why hydrogen, why so high T$_c$.}  The dearth of success of numerous high throughput
searches proposing higher T$_c$ materials, especially hydrides, is touched on briefly.
Based on as yet unapplied developments in simplifying effects of atomic displacement,
it is proposed that there is a straightforward path 
toward a deeper understanding of ``metallic hydrogen superconductivity" in conjunction
with added computational efficiency, and that 
some human-learning should assist in focusing the search for higher T$_c$ superconductors.  
\end{abstract}

\maketitle
%\onehalfspacing

\section{Introduction} 
The 2014-2015 discovery by Emerets' group\cite{Drozdov2014,Drozdov2015,Einaga2016} 
of superconductivity 
(SC) up to T$_c$=200 K in SH$_3$ in the 160-200 GPa range ignited a new era in SC research, and especially on understanding its microscopic origin and working toward even higher T$_c$ SCs. Remarkably, this compound had been predicted (independently and published earlier) in 2014 by 
Duan {\it et al.}\cite{Duan2014} using standard density functional theory (DFT) and 
Eliashberg formalism, to be a 200 K superconductor in the same pressure range. The following year saw the extension of the calculations and explanation of the
origin of very high T$_c$ by several 
groups.\cite{Ma2014,Papa2015,errea2015,bernstein2015,akashi2015,jose2016,Quan2016,errea2016}.

 This breakthrough was followed by the prediction and discovery of   
LaH$_{10}$ (T$_c$=250-260K around 200 GPa)\cite{LaH10theory,ME2018a,ME2018b,MS2019} 
and soon after YH$_9$ (243 K at 200 GPa\cite{Peng2017,Kong2021,Snider2021}) 
was predicted also before (or independent of) being confirmed 
by experimentalists, solidifying their origin 
as conventional, but now astounding, superconductors. Half a dozen or so compressed 
metal hydrides with T$_c$ above 100 K have been discovered since, displaying similarities 
and some differences.  

These metal hydrides are phonon-paired superconductors, as shown most dramatically by the
exceptionally large isotope effect\cite{Drozdov2015} in SH$_3$, which is what theory would predict.
After decades of development of the formal theory and then implementation into 
computational packages, the current state of prediction of new superconductors is 
on firm grounds.\cite{WEP2023} In spite of development and application of 
high-throughput searches (which necessarily require computational simplifications)
of binary and ternary hydrides with several predictions of 
new and possibly better examples, discovery of new examples has slowed. Quite possibly 
this limited success  is due to some lack of crucial insight into the origin of 
high T$_c$, rather than or in addition to the now common 
approach of (i) considering the full phase diagram, (ii) selecting by some
criteria the best suspects and checking for stability, then (iii) calculation of 
the Eliashberg spectral function $\alpha^2F$ and thereby T$_c$, 
that could provide focus to the search for better superconductors. This
numerically taxing approach is being addressed by various groups, often using machine
learning methods based on a few simple targets, with modest success so far. 
This paper presents the argument that there is one missing piece of the hydride puzzle 
that should provide human insight, improve computational efficiency, and provide 
some focus for computational searches.

\section{The conventional approach}
Conventional SC theory for elements is straightforward. Using density functional theory methods, 
calculate the electronic band energies $\varepsilon_{kn}$, determining the Fermi surfaces, and the 
phonon frequencies $\omega_{q\nu}$, on meshes through the Brillouin zone. For each phonon chosen, 
the potential change due to the phonon is calculated, then the electron phonon matrix elements 
(functions of $kn,k'n';q,\mu$) evaluated to integrate $\varepsilon_{k,n}$ and 
$\varepsilon_{k+q,n'}$ over the 
Fermi surfaces. One obtains the Eliashberg spectral function $\alpha^2F(\omega)$, from which the 
electron-phonon coupling (EPC) strength $\lambda$ is obtained for a single element SC 
as given by McMillan:\cite{McMillan1968}
\begin{eqnarray}
\lambda=\int \frac{2}{\omega}\alpha^2F(\omega) d\omega \rightarrow \frac{N(0) I^2}{M\omega_2^2}.
\label{eqn:lambda}
\end{eqnarray}
$F(\omega)$ is the phonon density of states, and $\alpha^2(\omega)$ is the squared EPC matrix element averaged
over all phonons of frequency $\omega$. 
The final expression is exact for elements, in which case the el-ion matrix elements 
$I^2$ hides the complication that arises for compounds; for elements it is 
\begin{eqnarray}
I^2=\frac {\sum_{k'n'}\sum_{kn}|<kn| \frac{dV}{d\vec R}|k'n'>|^2 
                                 \delta(\varepsilon_{kn})\delta(\varepsilon_{k'n'}) }
              {  \sum_{k'n'}\delta(\varepsilon_{k'n'}) \sum_{kn} \delta(\varepsilon_{kn}) },
\label{phononME}
\end{eqnarray}
where $V(r,\{\vec R\})$ is the electronic potential for ions at positions $\{\vec R\}$
and the derivative is respect to displacement of the atom at $\vec R$,
and the double average is over $k$ and $k'$ over the Fermi surface.

Binary and ternary compounds cause an essential complication. Each phonon involves 
displacement  of all atoms, and the preferred method is to calculate the kernel of the 
matrix element for each phonon, {\it i.e.} the potential change, to evaluate the
matrix element\cite{GiustinoRMP}
\begin{eqnarray}
g_{k,n;k+q,n';q\nu}= \langle k,n|\hat{\epsilon}_{q\nu} \cdot
                        \frac {dV}{d\hat{\epsilon}_{q,\nu}}|k+q,n'\rangle
\label{eqn:phononME}
\end{eqnarray}
for states $k$ and $k+q$ on the Fermi surface.
Each integrand of the matrix element involves displacements of all atoms involved 
in phonon $q,\nu$,
then the scalar product of the eigenvector with the gradient of potential, then
evaluation of the matrix element. The potential $V$ is the sum of the pseudopotential
and the density dependent potential $V^{DFT}$, to which we return below. The square
$|g_{k,n;k+q,n';q\nu}|^2$ occurs in $\alpha^2F$. 
There are so many integrands, {\it i.e.} matrix elements, that
they have not been studied, indeed it is unclear just how they should be analyzed.
These matrix elements are accumulated into $\alpha^2F(\omega)$ involving a sum over 
all phonons $\{q,\nu\}$, 
and over $k$ and $k+q$ restricted to near the Fermi surface. To begin to get to the point, 
each phonon displacement (hence, matrix element) involves contributions from 
every atom in the cell, and each phonon is calculated separately. It should
also be noted that quantities are no longer simply dependent on $N(0)$ but on
how the densities of states are distributed over the various atomic orbital
contributions around the Fermi surfaces.

The derivative is with respect to each phonon displacement eigenvector $\hat{\epsilon}_{q\nu}$,
and calculation of the matrix element is widely understood by the computational
community as costly. 
 The proliferation of indices, sufficient $q$-mesh grid (see comments in the
Appendix), and number of
atoms in the unit cell makes the computational task evident.  Decreasing 
each $q$-mesh separation by a factor of 2 leads to a corresponding factor of 8 in 3D, 
and increasing the number of 
atoms in the unit cell increases cost by a factor only experts should estimate. The bottom line 
is that the 11 atom cell in LaH$_{10}$ is a somewhat taxing endeavor, and increasing the number of atoms 
rapidly decreases enthusiasm for what might turn out to be an uninteresting result anyway.
Creative approximations are widely applied to alleviate the computational effort.

\section{Reviving the `enatom'}
There might be a more instructive way, and more efficient as well. 
Each `kernel' of the phonon matrix element 
 Eq.~(\ref{eqn:phononME}) involves 
the first order displacement of every atom in the unit cell. These responses
to phonon displacements are calculated 
over and over for every phonon $q,\nu$. 
This repetitive calculation (it will be shown) of each atom's linear response suggests looking 
for an alternative. Efficiency turns out to be a byproduct of the primary aim,
to encourage study and understanding of EPC. 

%\subsection{The enatom and its consequences}
Textbooks teach that separation of a solid's charge density into contributions from 
separate atoms is subjective, even arbitrary. While true as stated, six decades 
ago this statement was shown to address too specific a question, that is, too specific
a system (the static lattice). Considering a solid 
(let's say a crystal, though that is not necessary for the theorem), Ball demonstrated 
that such a separation is possible,\cite{Ball1,Ball2} with physical consequences.
His interest was in the density and consequences of atomic distortions. The interest
here is in the corresponding potential.

The displacement of a single atom $\vec R_j$ in 
a crystal with density $n(\vec r,\{\vec R\})$ depending on the atomic positions
deviating mildly from equilibrium positions $\{\vec R^{\circ}\}$ gives 
rise to a vector field, which 
can be decomposed into its irrotational and divergenceless fields as 
\begin{eqnarray}
\frac {\partial n(\vec r;\{\vec R\})} {\partial \vec R_j} 
             &=& -\nabla \rho_j(\vec r-\vec R_{j}^{\circ}) \nonumber \\
             & &    + \nabla \times \vec B_j(\vec r-\vec R_j^{\circ}).
\label{eqn:Ball1}
\end{eqnarray}
This expression is a purely mathematical statement: a vector field can be decomposed 
into the gradient  of a scalar function $\rho_j$ and the curl of a vector function 
$\vec B_j$, with each reasonably measured with respect to 
undisplaced position $R_j^{\circ}$ of the atom that was displaced. Ball showed, in a 
very simple demonstration (incorporating the infinitesimal displacement of the crystal as a
whole) with a profound result, that 
(1) the lattice sum of the scalars $\rho_j$ that move rigidly replicates the periodic density, 
giving a unique decomposition of the density into a sum of atomic contributions. 
In addition, (2) the lattice sum of the vector deformations (``backflows'') 
$\nabla\times\vec B_j$ vanishes.
%To emphasize: this result gives, to linear order in displacements, the 
%decomposition of the initial charge density into $\rho_j$ that moves rigidly, and ``backflow'' 
%terms $\vec B_j$ that lattice sum to zero for the reference density. 
This pair of quantities, both the rigid and backflow parts, were given the 
name\cite{jens2007} of ``enatom'' (`en' meaning in Greek `to cut 
from within') to distinguish it from several previous uses of `pseudoatom' for related 
but either weak pseudopotential or subjective decompositions, 
as related in an early paper.\cite{WEP1979}

Because it is central to the following description, this identical separation 
(which applies to any function of $\vec r$ that is parametrically dependent
on the atomic positions) is displayed here, the rigidly displaced pseudopotential
is straightforward.
Explicitly for the full DFT electronic potential $V^{DFT}(\vec r;\{\vec R\})$
(Hartree plus exchange-correlation): 
\begin{eqnarray}
\frac {\partial V^{DFT}(\vec r;\{\vec R\})} {\partial \vec R_j} 
          &=& -\nabla v_j(\vec r-\vec R_{j}^{\circ}) \nonumber \\
          & & + \nabla \times \vec W_j(\vec r-\vec R_j^{\circ}).
\label{eqn:Ball2}
\end{eqnarray}
This defines the rigid $v_j(r-R_j^{\circ})$ and backflow (or deformation) 
$\vec W_j(r-R_j^{\circ})$ fields of the 
first order change in potential. As for $n(r;{R})$, this rigid part 
is a precise decomposition of the total
potential $V^{DFT}$, and the lattice sum of undisplaced backflow parts 
$\nabla\times \vec W_j$ 
vanishes, written explicitly here for emphasis
\begin{eqnarray}
\sum_j v_j(\vec r-\vec R_j^{\circ})&=&V^{DFT}(\vec r;\{\vec R^{\circ}\}),\nonumber \\
\sum_j \nabla\times \vec W_j(\vec r-\vec R_j^{\circ})&=&0.
\end{eqnarray}
This derivative in Eq.~(\ref{eqn:Ball2}) is of course the gradient 
that appears in the electronic 
matrix element $g$ or $I$ along with the gradient
of the pseudopotential, which is available in current codes. 
Note that $\vec W_j$, like $\vec B_j$ above, is only defined to within a gauge, {\it i.e.}
the gradient of a scalar function, but $\nabla\times \vec W_j$ is unique (and physical)
and that is used in the EPC matrix elements. It in natural to discuss $\vec W_j$
in the divergenceless gauge.

This gradient, the matrix element kernel, has a limited range due to the strong local
screening in a metal. It can be obtained readily from DFT calculations in a supercell 
of reasonable size (viz. a 3$\times$ 3$\times$ 3 supercell). Either DFT perturbation
theory can be used to obtain the gradient (using formalism for an infinitesimal displacement) or from 
a few finite displacements on a reasonable mesh, which are relatively quick 
calculations even for 20+ atoms 
in the cell. If desired, and likely to be important for understanding, the Helmholtz 
construction,\cite{WEP1979} expressed simply in a Fourier expansion, can provide these 
rigid $v_j$ and backflow $\vec W_j$ 
components of the change in potential.  At this stage this appears to be only like a change in 
representation of the derivative, from phonon to individual atoms. For 
the compressed metal hydrides, it provides an essential 
simplification, as will be described.

\section{The small mass of hydrogen}
The foregoing separation of the potential holds for any solid, but these derivatives provide 
the kernel of the EPC matrix element. The importance feature discussed here relates 
to the mass difference between H and the metal(M), well recognized but not taken
full advantage of. 
Because of this mass difference $F(\omega)$ and $\alpha^2F(\omega)$
divide into two contributions separated by a gap, sometimes a substantial one. However, 
even for SH$_3$ with smaller mass difference than many (32:1), the contributions 
are divided into a low frequency spectrum that is, to excellent approximation, 
associated with the metal S, and a higher frequency region involving dominantly
H (this separation is not quite true for the lower pressures, at least in harmonic calculations.
This separation is much more the case for La in LaH$_{10}$), with its 139:1 mass ratio 
(because the heavy atom is unable to follow the rapid H motion). This observation is 
reminiscent of the ``double Born approximation'' of Onuorah {\it et al.}\cite{Onuorah}
for muons in solids ($m_{\mu}$=207$m_e$) -- electrons
respond almost adiabatically to the muon position. In hydrides, H responds roughly
adiabatically to the position of the heavy metal atom. Thus the separate H and
metal phonon eigenvectors form an approximately orthonormal eigensystem of the
respective vibrations, which becomes useful in the formalism.

Because $\alpha^2F$ separates, it is also true of $\lambda$ and the frequency 
moments,\cite{alldyn}
and $I^2$ is an atomic property (as is the mass), so it follows that  
\begin{eqnarray}
 \lambda=\lambda_M + \lambda_H;~~~~ \lambda_j=N_j(0) I_j^2/M_j \omega_{2,j}^2.
\label{eqn:separation}
\end{eqnarray} 
It is then only necessary to divide $N(0)$ into contributions from each atom, 
a somewhat subjective process but one that won't differ greatly amongst practitioners and should
be done such that the atomic contributions sum to $N(0)$.

This separation of $\lambda$ is broadly recognized, and the partial integral over
$2\alpha^2F(\omega)/\lambda\omega$ is standardly plotted as an integral over the integrand 
up to a limit of $\omega$,
providing $\lambda(\omega)$ that reveals the contribution from each regime, 
{\it i.e.} each atom.
Typically the metal contribution is 15-20\% of $\lambda$. This low frequency contribution 
is sometimes credited with 
providing the last increment to $\lambda$ that boosts T$_c$ significantly. 
Such is {\it not} the case, as described in the following section.

\begin{figure}[!ht]
%\centering
% \includegraphics[width=1.0\columnwidth]{./Isq-vs-P_H3S.png}
  \includegraphics[width=1.0\columnwidth]{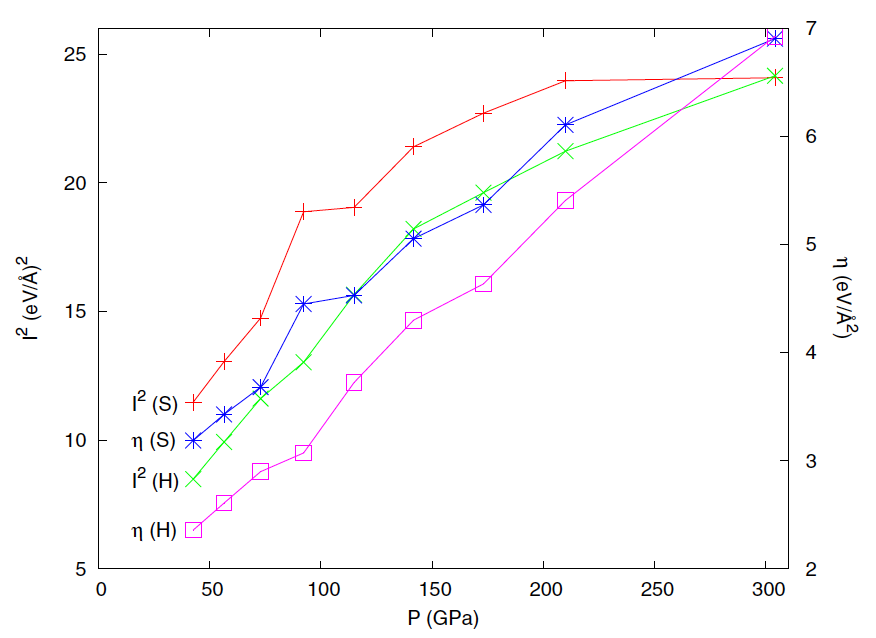}
 \caption{Pressure dependence of the separate S and H values for $I^2$ and $\eta=N(0)I^2$,
calculated from the Gaspari-Gy\"orffy multiple scattering expression.\cite{GG} 
The increase under
pressure is due almost entirely to the increase of $I^2$ (for each atom). The physical
regime for SH$_3$ is from 160 GPa and above, where the increase in $I_H^2$ is 25\%
by 300 GPa. Note that the calculated T$_c$ decreases for pressure above 200 GPa. 
The figure is reproduced from Ref.~\onlinecite{Papa2015}.}
 \label{fig:GG}
\end{figure}

The first study for hydrides of this separation was presented by Papaconstantopoulos
{\it et al.}\cite{Papa2015} who applied the rigid atomic sphere model 
(known  as the ``rigid muffin tin approximation'') of Gaspari and Gy\"orffy.\cite{GG}
This model was derived specifically to calculate the atomic $I_j^2$ for individual
atoms, and the model is very useful for
close packed transition metals. It becomes more approximate for superconductors where
the change in potential becomes less strongly local, or directional in character.

The result for SH$_3$ is shown in
Fig.~\ref{fig:GG}, along with $\eta_j=N(0)_jI_j^2$, for each of the two atoms.
The strong increase in $I_H^2$ (hence $\eta_H$) under pressure is surprising. After all,
increasing pressure, hence increasing density, might be expected to increase screening,
thus reducing the change in potential with motion and decreasing $I_H^2$. However, static screening
depends largely on $N(0)$.  Band broadening under pressure will decrease 
$N(0)$ and possibly screening, thereby enhancing $I^2$. However, it is the density
of states per unit volume that is comparable among materials, and the volume decreases
with pressure, giving some offset to the band broadening. 

These competing expectations require more
direct study. This calculated increase in $I^2$ (for both atoms) is dominant 
in SH$_3$, so the expectation that H, without any core to screen, should be a
stronger scatterer than typical metals (with substantial cores, hence effective
pseudopotentials), seems not to be the case.  However this superconductor
is not the best to judge because $E_F$ lies very near the peak of a sharp 
and narrow van Hove singularity,\cite{Ghosh2019} (vHs) where behavior under pressure 
is uncertain to estimate. Also, additional physics arises, including electron
velocities vanishing (leading to non-adiabatic effects) and the electron 
scattering by phonons $\varepsilon_k+\omega_q\rightarrow\varepsilon_{k+Q}$ involves
scattering across the vHs, {\it i.e.} the density of states N($\varepsilon$) as
well as the velocities results in more complex behavior.

\section{Metal and Hydrogen Atoms Separately}
The focus therefore turns to hydrogen specifically
\begin{eqnarray}
T_c=F(\omega_{log,H},\omega_{2,H},\lambda_H);~\lambda_H = \frac{N_H(0)I_H^2}{M_H\omega_{H,2}^2},
\end{eqnarray}
with analogous expressions for the metal with $M$ subscripts (which we will be ignoring). 
With the other factors being calculated (with the known mass) the property that is unknown is
$I_H^2$, and it has attracted almost no attention in hydrides beyond that mentioned above. 
The M-H separation of $\lambda$ and the frequency moments allows a conceptual exercise: 
what would T$_c$ be if only the M, or only H, 
contributions were present? Note that the sum of these would not add to the true T$_c$
due to the many nonlinearities that are involved, but the sum is not of interest. 

\begin{figure}[!ht]
%\centering
% \includegraphics[width=0.5\columnwidth]{./Isq5-vs-P.png}
  \includegraphics[width=0.5\columnwidth]{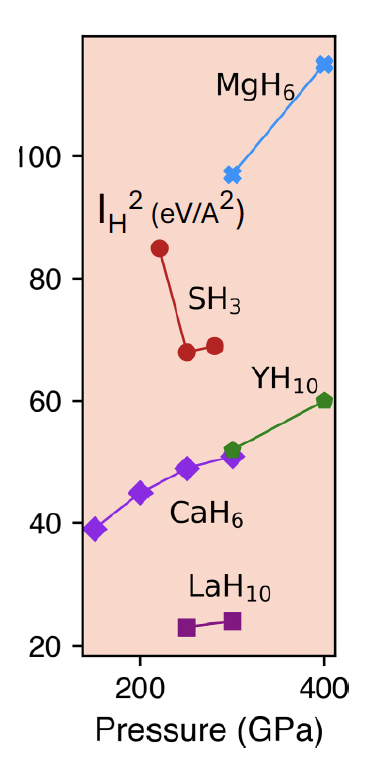}
  \caption{Pressure dependence of $I_H^2$ (hydrogen alone) for five metal hydrides at
assorted values of pressure, extracted from Eq.~\ref{eqn:separation} when all the other
factors are calculated. The factor of five spread is unexpected, and the factor of two
between CaH$_6$ and isovalent, isostructural MgH$_6$ at 300 GPa raises questions. The 
highest T$_c$ metal hydride at this time, LaH$_{10}$, has the lowest $I_H^2$ of
this group, without any explanation yet. 
 The figure is reproduced from Ref.~[\onlinecite{Quan2019}].} 
 \label{fig:Isq5}
\end{figure}

Quan {\it et al.}\cite{Quan2019} considered five compressed metal hydrides -- SH$_3$,
LaH$_{10}$, YH$_{10}$, CaH$_6$, MgH$_6$ -- each at a few different pressures, and
studied the separate metal and hydride contributions to T$_c$. The results for $I^2_H$
(the other quantities are calculated or known) are  presented in 
Fig.~\ref{fig:Isq5}. To summarize: when the metal contribution to
$\alpha^2F$ was discarded, T$_c$ was {\it essentially unchanged} (and the tiny
contribution to T$_c$ could be negative as well as positive). 
The decrease in $\lambda$ by ignoring the M contribution, which 
decreases T$_c$, was canceled by the increase in $\omega_{log}$ and $\omega_2$, 
which increase T$_c$. 
 When the H contribution
to $\alpha^2F$ was deleted, of course the metal atom $\lambda_M\sim 0.15-0.25$ on its own
gives no superconductivity. This picture aligns well with the original
concept of Ashcroft\cite{Ashcroft1} -- metallic hydrogen -- that high H 
frequencies and moderate screening would lead to high values of T$_c$. 
Many numerical values of related quantities are given in the tables in 
Ref.~[\onlinecite{Quan2019}]. 

The conclusion is that ignoring the M contribution does not affect
T$_c$, {\it i.e.} these metal hydrides are simply {\it metallic hydrogen superconductors},
with the metal contribution providing stability of the compound but confusing
the source of superconductivity.  This picture is also
close to the second vision of Ashcroft,\cite{Ashcroft2} using H-containing molecules to
provide pre-compressed reservoirs of hydrogen and resulting hydrogen superconductivity.
So far high symmetry structures have provided the successes, that is a separate subject.
Turning attention now specifically to the H parameters should
enable essential quantitative input into the high values of T$_c$.
The attractive feature of computational efficiency will be discussed in the next section.

The results of Quan {\it et al.}\cite{Quan2019} give a sharp indication of the next level
of questions to be addressed. The derived values of $I_H^2$ versus pressure,
presented in Fig.~\ref{fig:Isq5} but for a model of scattering,\cite{GG} is
startling in a few respects. The dominant trend is toward higher
values as pressure increases, as for the GG values\cite{GG} for SH$_3$.
Unexpected is that the strengths of scattering $I_H^2$ differ considerably
amongst the five hydrides, especially considering that SH$_3$ and LaH$_{10}$ 
are the first two confirmed high T$_c$ members, both having T$_c$ of 200K or greater. 
Overall the values differ by a factor
of five over the range 200-400 GPa for the cases that were calculated. 

It is noteworthy that the highest T$_c$ hydride LaH$_{10}$ `scored low'
on this competition, very surprising considering its highest confirmed
T$_c$$\sim$250-260 K.
Another surprise is the factor of two difference between CaH$_6$ and MgH$_6$, which
needs an explanation, similarly for LaH$_{10}$ and YH$_{10}$. 
A path forward that might provide the explanation is the implementation and
analysis of the enatom potential and matrix elements. The
non-monotonic behavior of SH$_3$ is likely because E$_F$ lies
very near the top of a van Hove singularity,\cite{Ghosh2019} 
which makes pressure effects delicate
[especially N(0)] and requires additional theory (the phonons scatter electrons
from a region below E$_F$, below the peak, to above E$_F$, above the peak, 
requiring corrections to the 
usual constant $N(E)\approx N(0)$ treatment) and computational accuracy for 
accurate predictions (an example is given in the Appendix).

\section{Calculation: methods and efficiency}
\subsection{$q$-mesh convergence}
The formalism, calculational results, and derived focus on H lead to questions of,
first, a clearer understanding of this `hydride superconductivity,' and second,
computational methods and efficiency. The electronic band eigensystem is always required,
as is the phonon eigensystem (frequencies and eigenvectors). 
The currently prominent code is EPW\cite{EPW2}, 
which provides the linear response evaluation of $N_q \times N_{\nu}$ 
derivatives, one for each $q$ point on a $N_q$ mesh and each branch $\nu,$
each involving all atoms in the cell.  For a relatively hard case of
LaH$_{10}$, a 4$^3$ $q$-mesh (giving very questionable accuracy) would leave of the 
order of $\approx$10$\times$33 branches $\sim$ 330 linear response calculations 
(cubic symmetry if fully implemented\cite{EPW2} would reduce the 4$^3$ factor 
to no more than 10 $q$ points in the irreducible Brillouin zone). A 6$^3$ mesh would
increase this to around 1000 linear response calculations, depending on symmetry. These
estimates, if reasonable, explain why 8$^3$ $q$~meshes or higher have been rare. 

The Appendix contains one example, a seemingly simple case, where convergence with respect 
to $k$- and $q$-meshes
are unexpectedly difficult to achieve. For that simple Fermi surface (lithium
under pressure), it is uncertain
how well converged $\alpha^2F$ and everything that is obtained from it is, even for
a 48$^3$ $k$-mesh and 24$^{3}$ $q$ mesh. The issue is that large 
contributions arise from strongly localized nesting regions across the Fermi 
surface, with the effect of having $q$-dependence of coupling strength on arcs
or lines across the zone.
Similar effects should be more prevalent for complex, multisheeted Fermi surfaces.

\subsection{Symmetry and efficiency}
Considering the real space focus on H now, in SH$_3$ the tetragonal site symmetry of H
leaves only two linear response calculations, one longitudinal (displacement of H toward a
neighboring S atom) and one perpendicular, plus use of tetragonal symmetry\cite{EPW2}
of the H site. If the sulfur value of $I_S^2$ is wanted (and it might be instructive),
its cubic symmetry would require only one displacement calculation, plus symmetry. 
For LaH$_{10}$, there are
two H sites, each with some symmetry, so 4-5 linear response calculations would be required.
The phonon eigenvectors are already computed, so construction of
the phonon potential 
$\hat{\epsilon}_{q\nu}\cdot\partial V/\partial \hat{\epsilon}_{q\nu}$ is a linear algebra 
operation. With the electronic wavefunctions in hand, the remaining evaluation of the 
matrix element is also a linear algebra evaluation. The numerical speedup
should be quite significant, at the expense of some additional coding. 
The results for $\partial V/\partial \vec R$ and its rigid and deformation parts should 
provide information
beyond $I_H^2$ about the enatom potential of the H atom in different environments. Does the
deformation potential ever become important? is one question that comes to mind.

\section{Previous Indication from an Enatom}
\subsection{Lithium as a high T$_c$ superconductor}
The title of this paper referred to understanding rather than any computational efficiency.
Metal hydride superconductivity increased from 10 K (PdH) at ambient to 250-260 K around
200 GPa in LaH$_{10}$, a factor of 25. There is a related but yet more extreme example. 
Lithium, finally
discovered to be a 180 $\mu$K superconductor\cite{Tourin} in 2007, when squeezed to 35 GPa
achieves $\sim$20 K superconductivity - around five decades of 
increase.\cite{Shimizu2002,Struzhkin2002,Deemyad2003} 
This increase,, to the highest T$_c$ for any element at the time, 
is almost unbelievable for a weak pseudopotential, nearly free electron monovalent 
metal. The story is an instructive one, see the original papers for the
full story.\cite{Deepa2006,Deepa2007} 

Pressure transforms
bcc Li to fcc Li, upon which the spherical bcc Fermi surface develops necks across the 
L point (very much like the monovalent Cu Fermi surface) where the electron velocity 
drops by more than 
a factor of two, and strong and very localized 
nesting features\cite{Gilat,WEP1975,WEP-TCM}  
arise, resulting in unexpected, and strong  `fine structure' in the $q$ dependence of
the phonon linewidths (nesting function with electron-phonon matrix elements inserted)
and their associated contribution to $\alpha^2F$ and T$_c$. Not a large N(0), but an 
$I^2$, and $\alpha^2$, that was difficult to converge even with very fine 
$k$- and $q$-meshes, numbers and a figure are given in the Appendix. More complex
Fermi surfaces will enhance such occurrences, but Li shows it can happen for even the
simplest of Fermi surfaces.  

\subsection{Glimpse of the Li enatom}
Around the same time the enatom was calculated, as a demonstration of the method, 
by Kunstmann {\it et al.}\cite{jens2007,JKthesis} 
for the simple metals Li  and 
Al (T$_c$=1.2K) at ambient, in part because they
could be represented by local pseudopotentials and also because they display different
behavior under pressure.  The very large T$_c$ increase in Li is described above, 
making it the highest T$_c$ element at the time.
For Al, T$_c$ decreases under pressure, with unreachable temperatures above 20 GPa. 

Not surprisingly for these nearly free electron metals (except beyond expectation one could 
display high temperature superconductivity), the rigid
potentials, restricted by symmetry to be cubic, were found to remain close to
spherical under pressure. (For Li the rigid density included a {\it negative value} in the
region of second and third neighbors, such behavior is allowed.)
The non-spherical components of the rigid potential could be identified,
 and also the  Friedel oscillations in the rigid potential. With no
surprise, the backflow potentials, restricted to have the cubic symmetry of a vector
field and whose lattice sum must vanish, seemed (without much to compare to) to be small.
Seemingly, matrix elements (not computed) would be dominated by the displacements 
of the spherical rigid potentials $v_j$ and pseudopotential, with little effect from
the deformation potential. A glimpse of the deformation density and potential of
Li at 35 GPa is given in the Appendix.

\begin{comment}
The increase in T$_c$ to 20 K in Li
could be attributed to the appearance of $2p$ character in the
electronic structure, as the Fermi surface evolves from spherical to extended Cu-like. 
The combined electronic structure and phonon character led to ``hot spots''
on the Fermi surface involving nesting of Fermi surface scattering combined
with larger matrix elements. The original papers should
be consulted for more details.\cite{Deepa2006,Deepa2007}
General extension of similar studies to $s, p$, and $d$ orbitals should be
straightforward; derivatives of non-local pseudopotentials are available, 
and the double Fermi 
surface average (without matrix elements known as the nesting function) was implemented 
nearly five decades ago.\cite{WEP1975} 
\end{comment}

\section{Searching for higher T$_c$}

We now revert to a tertiary purpose of this paper (beyond understanding and
efficiency): to contribute to the search, in
progress by many groups, to discover and predict competently even higher T$_c$
hydrides under pressure, but ideally at ambient conditions. A given higher T$_c$
hunt might have an emphasis on larger $\lambda$, keeping in mind that this direction invites
other instabilities. Pushing strong coupling to higher frequencies might
be a different goal. These two figures of merit for T$_c$ -- $\lambda$ and a
frequency moment $<$$\omega$$>$ -- have dominated the discussion since the time of
McMillan's Eq.~\ref{eqn:lambda}. 

\subsection{A single figure  of merit}
This tradeoff was addressed five decades ago 
by Leavens and Carbotte (LC),\cite{Leavens1974,Leavens1977} who were surveying the 
strong coupling materials of the 1970s in terms of Eliashberg theory. 
As a different measure of the underlying influence of
$\alpha^2F(\omega)$, which is what determines T$_c$, they found that
a different focus rather than $\lambda$ and  $(2/\lambda)\alpha^2(F(\omega)/\omega$ 
showed promise.
Their focus was on the area ${\cal A}$ under $\alpha^2F(\omega)$, rather 
than moments of $\alpha^2F(\omega)/\omega$,
gave a simple and impressively good fit to the experimental data at the 
time. Their fit to data gave (temperature and frequency will be expressed in
the same energy units, {\it i.e.} $k_B$=1=$\hbar$)  
\begin{eqnarray}
 T_c^{LC} = 0.148 {\cal A} = 0.074~\omega_1 \lambda
\end{eqnarray}
as a good representation of T$_c$ (the constant
would depend somewhat on $\mu^*$, the last expression uses the moments of
Allen and Dynes)). One interpretation of ${\cal A}$ is
that this area is the product of the average of $\alpha^2F$ over the 
full interval [$0,\omega_{max}$] times the length $\omega_{max}$ of the interval.
Roughly speaking, $\lambda$ is the measure of the strength of $\alpha^2F$, while
$\omega_1$, which is independent of the strength of $\alpha^2F$ but depends
only on its shape and extend, is the measure of the spread of the coupling.
The distribution over frequencies in this interval would play no 
discernible part. Restating, the low frequency part
$\lambda$ would be balanced by the higher frequency measure $\omega_1$. 
One implication is that increasing one while keeping the other constant would 
increase T$_c$, similar to the usual picture but differing in numerical detail.
Since $M\omega_2^2$ is independent of mass for a harmonic elemental solid, 
the LC expression has an $M^{-1/2}$ mass dependence, following the standard
isotope shift for elemental SCs.

\subsection{Application to hydrides}
\begin{figure}[!ht]
  \centering
  \includegraphics[width=1.00\columnwidth]{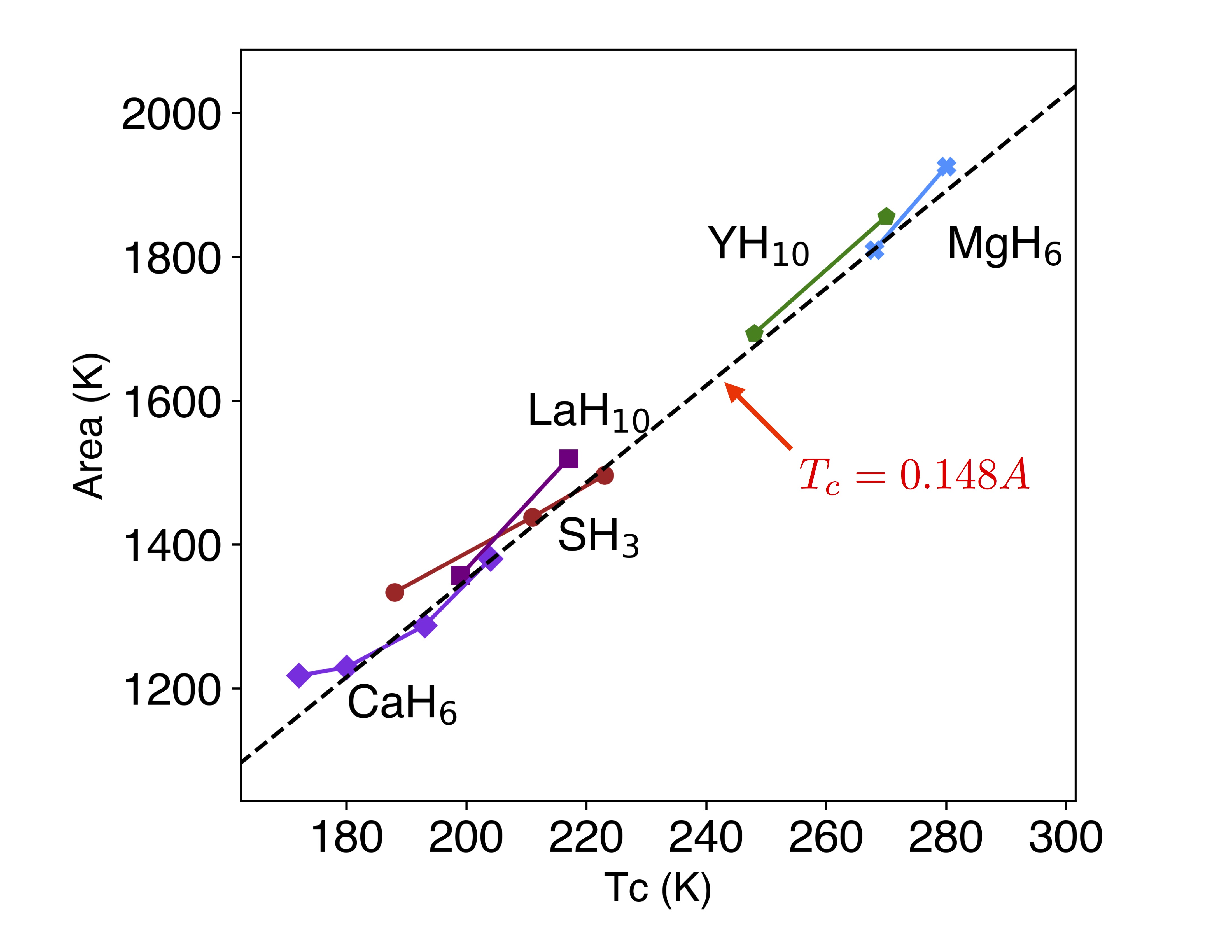}
\caption{A scatterplot of the area ${\cal A}$ versus T$_c$
for the five binary hydrides at the various pressures that were
calculated by Quan and Pickett.\cite{Quan2019}
Note that T$_c$ and ${\cal A}$ are calculated from the H vibrational spectrum of
$\alpha^2F$.  The strong correlation is evident. The slope of  0.148 denotes
the Leavens-Carbotte line for strong coupled intermetallics existing in 1974.}
\label{fig:Leavens}
\end{figure}

One can then ask whether this ${\cal A}$ figure of merit has relevance in compressed hydrides. 
The correlation between ${\cal A}$ and T$_c$ for the five compressed hydrides shown in
Fig.~\ref{fig:Isq5} is provided in Fig.~\ref{fig:Leavens}. Recall that only the 
H part of $\alpha^2F$ is used in the results in both figures, the metal hydrides have
become more simply ``hydride superconductors.''. The excellent correlation is evident. 
Moreover, the proportionality (slope of the line) is the same as for the 1970s elemental
superconductors and 21st century compressed metal hydrides (once the metal
contribution is neglected). This equivalence in value
and slope is is remarkable: the hydrides have T$_c$ up a factor of 25 or more higher,
``because'' the area ${\cal A}$ is that much larger. 
but it also indicates that T$_c$ for compressed metal hydrides is far from optimal:
$N(0)$ or/and $I_H^2$ need to be increased for higher T$_c$.

The importance of $N(0)$ has always been forefront, the importance of $I^2$
instead of frequencies now becomes forefront (in truth it always was).
Coupling throughout the frequency spectrum of $\alpha^2F$, with amplitude as 
large as possible and frequencies extending 
as high as possible, seems to be the key to high T$_c$ in compressed metal hydrides
as well as in the 1970s high T$_c$ materials. 
Likely these results apply to ambient condition hydrides as well, 
which are currently being actively sought with little success so far in the 
synthesis of predicted candidates. 

The LC expression T$_c$ versus ${\cal A}$  is somewhat reminiscent of the strong coupling
limit of T$_c$ within Eliashberg theory established by 
Allen and Dynes (AD)\cite{alldyn}, obtained formally as
\begin{eqnarray}
 T_c^{max,AD}=0.18~\omega_2\sqrt{\lambda}=0.18\sqrt{\frac{N(0)I^2}{M}}.
\end{eqnarray}
The LC expression can be written (for an element)
\begin{eqnarray}
T_c^{LC}=0.07 \frac{\omega_1}{\omega_2}\frac{N(0)I^2}{M\omega_2},
\end{eqnarray}
thus scaling differently with $\eta=N(0)I^2$ but with the same (correct) scaling with mass. 
The LC expression can be regarded as a linearization of $T_c(\lambda)$ in a regime that
is still far from the AD limiting regime. This view would then suggest that 
current hydrides are still well below the limiting regime of T$_c$, as
indeed the AD equation for T$_c$ indicates.

\subsection{A differential viewpoint}
To repeat: the LC relation indicates that it is the mean magnitude of $\alpha^2F$ over 
as large a range as possible that determines ${\cal A}$ and thus T$_c$, and not
an explicit frequency distribution.  Can this (phenomenological) view be
reconciled with the rigorous result of Bergmann and Rainer (BR) from Eliashberg
theory that suggests a different viewpoint? (BR)\cite{Rainer} BR calculated the 
functional derivative 
\begin{eqnarray}
{\cal D}(\omega)\equiv \frac{ \delta T_c[\alpha^2F,\mu^*]}{\delta\alpha^2F(\omega)},
\end{eqnarray} 
which gives the
increase in T$_c$ that will result from an extra increment $\Delta\alpha^2F$
at $\omega$. This is a material dependent function, {\it i.e.} dependent on
$\alpha^2F$ but finally reflecting little if any  real dependence on material.  This
function is linear at low $\omega$ (indicating a poor choice of frequency region
for increasing T$_c$), it peaks at a value $\delta_m$, just above $\omega\sim 2\pi$T$_c$
(more precisely, around 6.5$T_c$), then decreases rather slowly beyond. Thus if 
strength can be increased at high frequency it is not so important at
what frequency, but it is better than adding at low frequency, the separation
occurring at $\delta_m$. 

The resolution to this difference is (at least) twofold. First, 
the rigorous results of  BR for the derivative 
(essentially the same for several elemental SCs when scaled by T$_c$) 
were for strong coupled SCs with T$_c$ up to 15K, 
mostly heavy metals. Different classes, with different 
distributions of $\alpha^2F(\omega)$, might show different behavior. Second  
is that one usually cannot simply add new phonons with coupling to a material. 
However, one can imagine 
altering a material to give a shifts in $\alpha^2F$ weight in  specific
regions of frequency, viz. by substituting (similar) atoms, or by external means 
(strains, pressure, or boundaries).

This question was elaborated by Camargo-Mart\'inez {\it et al.}, who evaluated 
$\alpha^2F$ and ${\cal D}(\omega)$ for SH$_3$ at pressures from 215 GPa 
down to 155 GPa, where their (harmonic) T$_c$ values increased from 138K to 203K,
correlating reasonably with the experimental data. Their peak position $\delta_m$ 
of ${\cal D}(\omega)$ occurred near 7$k_BT_c$ (versus 6.5$k_BT_c$ for BR),
but otherwise showed the same frequency dependence.
SCs. Upon lowering pressure, the lower frequency peak of  H vibrations
shifted downward and coincided with $\delta_M$
at 155 GPa, and the shift in T$_c$ correlated well with the shift in ${\cal D}(\omega)$
as pressure was lowered. Comparing values of ${\cal A}$
is not possible (without certain projections) because the S and lower H vibrations become mixed at
the lower pressures, in spite of the factor of 32 difference in atomic masses.

\section{Summary}
The comment in the title ``why compressed hydrides are such high T$_c$ superconductors''
has yet to be resolved. An approach to enable answering this question
has been outlined here. This viewpoint involves (i) recognizing and applying the enatom picture,
(ii) enabling the enatom potential, which also promises more  efficient calculations,
that (iii) will provide more insight (human learning) into hydride electron-phonon coupling.
Calculation of scattering from $\vec k$ to $\vec k + \vec q$, both on the Fermi surface, will
provide the required information. The study of Li under pressure\cite{Deepa2006,Deepa2007}
identified strong coupling between very specific hot lines across the Fermi surface.  
Such ``details'' are likely to arise more often in multiband Fermi surfaces,
including compressed metal hydrides, and may require find $q$-meshes to obtain converged values
of superconducting parameters. For Li under pressure, the rigid potential shift is
dominant in the electron-photon 
matrix element. What is unclear is whether it is the phonon perturbation $dV/dR$ itself,
or instead Fermi surface properties (wavefunction character, nesting, etc.) that 
enter the matrix elements that are the
underlying origin of the amazingly high temperature hydride superconductivity. Well converged
calculations are required to resolve these possibilities.

The enatom picture has other applications involving phonon- or strain-related transport. 
One example is the response of electronic
states and band energies to strains or pressure. Khan and Allen proved,\cite{Khan1984} 
after a few decades of study by prominent theorists, that the strain deformation potential
-- shift in $\varepsilon_k$ due to a strain -- is given by the band diagonal, $q$=0,
matrix elements of the enatom potential appended by a simple kinematic quantity.
Likely there are other applications of the enatom potential and associated density
in the many manifestations of transport theory.

%%%%%%%%%%%%%%%%%%%%%%%%%%%%%%%%

\section{Acknowledgments}
This contribution was provided in memory of Mikhail Eremets, the leader of the 
high pressure group
at MPI Mainz that discovered the 200K superconductor SH$_3$ (and LaH$_{10}$ afterwards). 
Mikhail was a leader in the field of high pressure physics, 
a thoughtful coauthor\cite{physicstoday}, a conscientious guide
for lab tours (twice) for this author, a valued friend, and an exemplary member
of his scientific community. A visit to Liverpool hosted by Michael Ball 
ca. 1976 confirmed
my decision to keep this `enatom' picture in mind. My coauthors on earlier work  
on the enatom of Li and Al,
Jens Kunstmann and Lilia Boeri, deserve acknowledgment for their collaboration
in the midst of their primary research activities. I also acknowledge the 
construction of a picture of coupling in Li, perhaps a resolution but not 
quite of publication
convincability, by Jan Kune\v s, of the superconductivity of Li resulting 
from a convergence 
of electron $2p$ character of matrix elements coinciding with strong nesting
regions on the Fermi surface. 
Input from Christoph Heil has proven helpful. 

\appendix
\section{Example from Lithium under Pressure}
\subsection{The enatom deformation potential}
While the enatom picture has not yet been implemented in electron-phonon codes,
there is one instructive example. Lithium, whose T$_c$ increases around five decades 
of temperature from ambient to 35 GPa (sub-mK to 20K),
was discussed in the text and referenced.  Using finite difference calculations
the enatom density and potential were straightforward to obtain.\cite{jens2007,JKthesis}

\begin{figure*}[!ht]
  \centering
   \includegraphics[width=2.00\columnwidth]{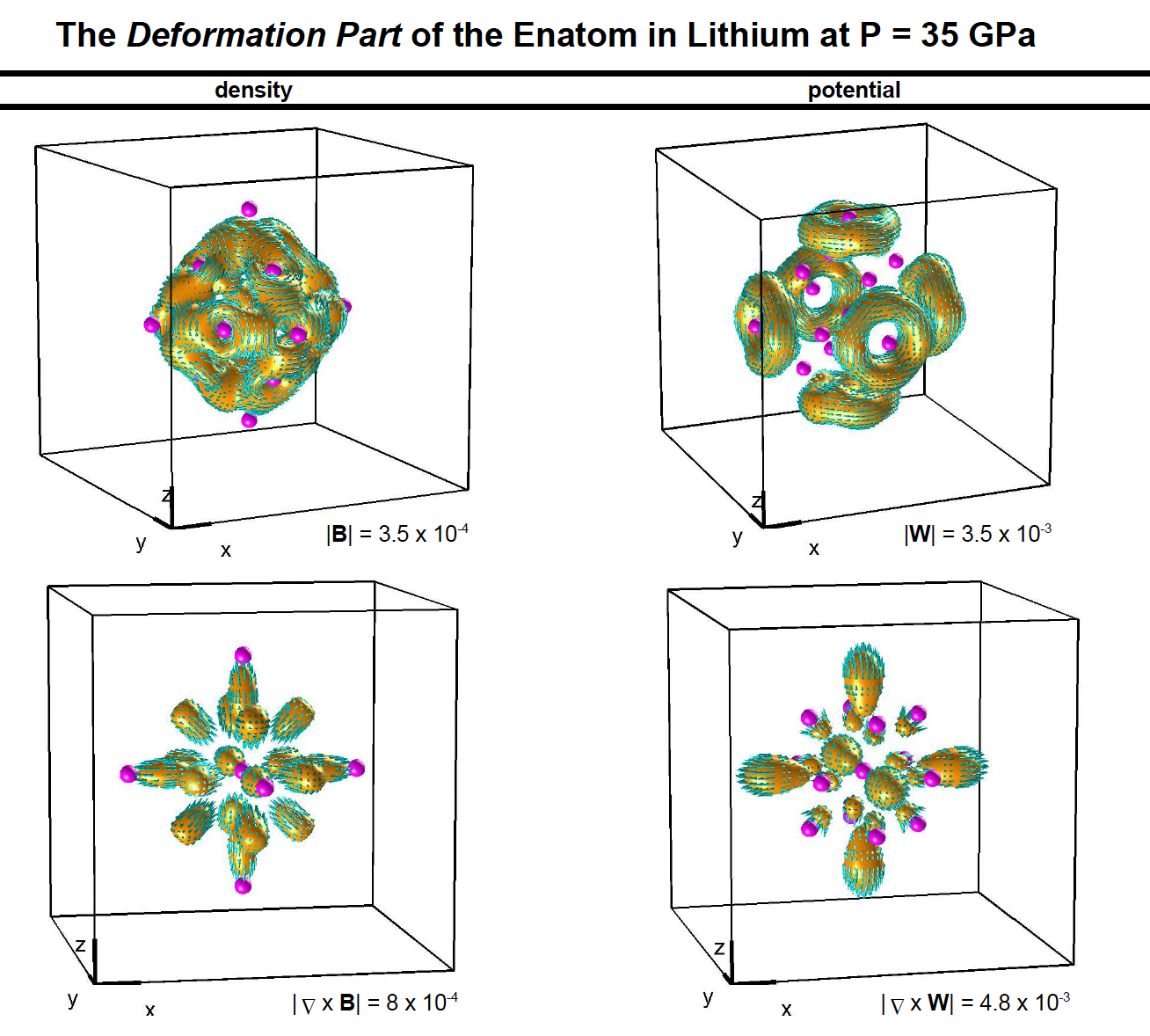}
  \caption{Isosurface plots of the enatom deformation density and potential quantities:
gauge dependent $\vec B$ and $\vec W$, respectively (divergenceless gauge was
chosen), and the physical quantities 
$\nabla\times \vec B$ and $\nabla\times \vec W$, respectively, 
for fcc lithium at 35 GPa. The directions of
these vector quantities are denoted by arrows on the isosurfaces of their
magnitude. $\vec B$ might be considered as a ``beating heart'' isosurface, with 
rather involved direction dependence on this surface but obeying the cubic
symmetry of a cubic field. The rigid quantity $\vec W$ appears as swirling
donuts near surrounding Li atoms in cubic symmetry form.
Note that when the 
bottom two curl products are dotted into a displacement in the $+\hat{x}$ direction,
the effect is to move density/potential from in front to behind the displaced
atom  -- the deformation effect. These deformation effects appear to be small 
for Li, but no matrix elements have been calculated to allow assessment of their effect.
Credit goes to J. Kunstmann\cite{JKthesis} for calculation and
arrangement of the panels.}
  \label{Li_35GPa}
\end{figure*}

Figure~\ref{Li_35GPa} provides insight into the behavior of the potential deformation
quantities (as labeled) of Li under displacement,
$p$ and $d$ electron materials will display more complex behavior. The calculation
was done in a (3$a$)$^3$ 108 atom fcc supercell (using the structure at the volume
of the experimental pressure). The four panels picture the deformation density and potential
character, and are described in the caption. A separate interesting result (not shown) is
that the enatom rigid density contains a sphere shell of negative density around
and somewhat beyond nearest neighbors (which would serve to make the lattice
sum equal to the crystal density).

\subsection{Delicacy of electron-phonon coupling}
\begin{figure*}[!ht]
  \centering
   \includegraphics[width=1.20\columnwidth]{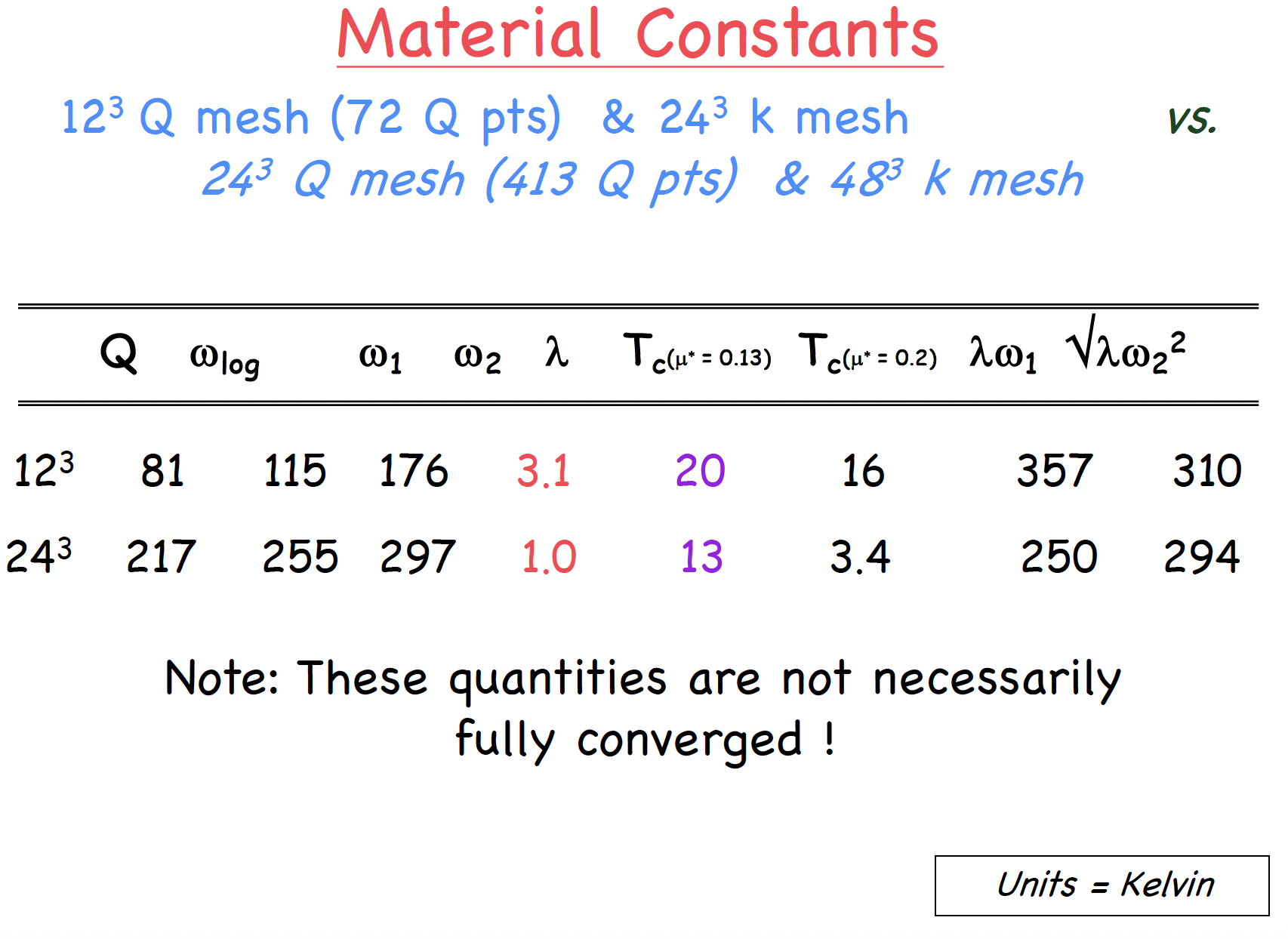}
  \caption{A slide from a presentation on the calculation of Li superconductivity
at 35 GPa.\cite{Deepa2006,Deepa2007} Even at quite fine $k$- and $q$-meshes
on a very simple Fermi surface, halving the mesh distance
gives extremely different values of critical parameters, see $\lambda$ and
the frequency moments. 
The large uncertainties in $\lambda$ and $\omega_{log}$ tend to offset in the 
resulting T$_c$, but leave confusion about what parameter(s) are important. 
(The strong effect of $\mu^*$ is evident as well.) The
fine $q$-mesh is necessary to see in the phonon dispersion curves where the Kohn
anomalies (large and sharply defined renormalization regions) 
occur.\cite{Deepa2006,Deepa2007} 
}
  \label{Li_parameters}
\end{figure*}

Figure~\ref{Li_parameters} provides the values obtained for several parameters from
calculations of 12$^3$ and 24$^3$ $q$-meshes, with $k$-meshes a factor of 2 finer
in each of the three directions. These data reflect the large
differences in $\alpha^2F$ that result for fcc Li at 35 GPa, even for fine $k$-
and $q$-meshes almost never used for hydride superconductors. These difference are
themselves the result of very small regions in the BZ with large nesting of the
Fermi surfaces. As stated, even this fine mesh may not give the converged values
of the various material properties. The calculations were done with Savrasov's
linear muffin-tin orbital code.\cite{Savrasov1,Savrasov2} T$_c$ was calculated from the 
Allen-Dynes equation.\cite{alldyn} 

\begin{figure}[!ht]
  \centering
   \includegraphics[width=1.00\columnwidth]{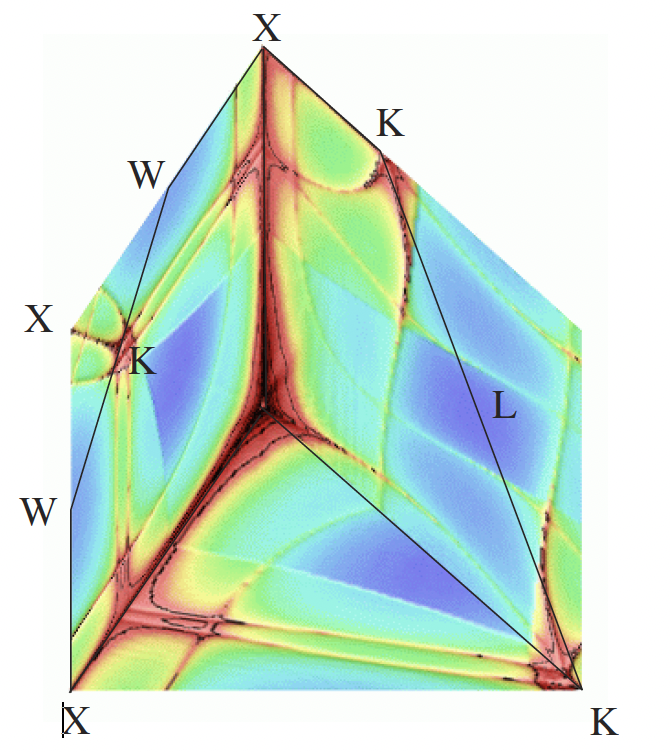}
  \caption{A plot of the nesting function of fcc Li at 35 GPa where it is
superconducting near 20 K,, showing intensity plots throughout three
of the high symmetry planes of the zone.\cite{Deepa2006,Deepa2007} 
Red indicates high intensity, deep blue is very low intensity.
The deep red area at the center of the plot, at the $\Gamma$ point, is a trivial
and non-physical divergence of the nesting function. The narrow stripes of this
nesting function are indicative of numerical difficulty in obtaining the volume integral,
which relates (when matrix elements are included) to $\alpha^2F$, 
coupling strength $\lambda$, phonon
moments, and T$_c$. 
}
  \label{Li_Nesting}
\end{figure}

The intricacy of the nesting function in fcc Li is displayed in Fig.~\ref{Li_Nesting}.
In spite of the very simple Fermi surface\cite{Deepa2006} the nesting function shows
high intensity along narrow sheets through $\Gamma-X$ (near $X$) and around the $K$ point
of the zone. The phonon linewidth, {\it i.e.} the strength of electron-phonon coupling,
is the nesting function with squared electron-phonon matrix elements inserted into
the integrand. Matrix elements
will reduce the $\Gamma$ point divergence to finite
values of the coupling. Also, the small weight given to the $\Gamma$ point 
(a single point in a 3D zone) in the
integral usually does not leave much contribution to the net coupling, if the
double $\delta$-function is taken care of properly. The point is that even simple
Fermi surfaces can have ``hot spots'' that make for slow convergence of integrals
such as those for $\alpha^2F$ and $\lambda$. There are correspondingly sharp Kohn
anomalies in the phonon spectrum.\cite{Deepa2006,Deepa2007}

\end{document}